\documentclass[preprint,3p,twocolumn,authoryear]{elsarticle}
\usepackage{amsfonts} 
\usepackage{amsmath}
\usepackage{amssymb}
\usepackage{graphicx}
\usepackage{subfigure}
\usepackage{color}
\begin{document}
\begin{frontmatter}
\title{Large amplitude electromagnetic solitons in a fully relativistic magnetized electron-positron-pair plasma}
\author[GB]{Gadadhar Banerjee\corref{altaffi}}
\ead{gban.iitkgp@gmail.com}
\cortext[altaffi]{On leave from Department of Basic Science and Humanities, University of Engineering \& Management (UEM), Kolkata-700 160.}
\address[GB]{Department of Mathematics, Siksha Bhavana, Visva-Bharati (A Central University), Santiniketan-731 235, India}
\author[SD1,SD2]{Sayantan Dutta}
\ead{dutta@irsamc.ups-tlse.fr; sayantan.dutta110@gmail.com}
\address[SD1]{Laboratoire de Physique Théorique, IRSAMC, UMR CNRS 5152, Université Paul Sabatier, 31062 Toulouse, France }
\address[SD2]{Institut de Recherche en Informatique de Toulouse, UMR CNRS 5505, Université de Toulouse, 31062 Toulouse, France}
\author[GB]{A. P. Misra\corref{cor1}}
\ead{apmisra@visva-bharati.ac.in; apmisra@gmail.com}
\cortext[cor1]{Corresponding author}
\begin{abstract}
Nonlinear propagation of purely stationary large amplitude electromagnetic (EM) solitary waves in a magnetized  electron-positron (EP) plasma is studied using a fully relativistic two-fluid hydrodynamic model which accounts for  physical regimes of both weakly relativistic  $(P\ll nmc^2)$ and ultrarelativistic $(P\gg nmc^2)$ random thermal energies. Here, $P$ is the thermal pressure, $n$ the number density and $m$ the mass of a particle, and $c$ is the speed of light in vacuum.
  While both  the sub-Alfv{\'e}nic and  super-Alfv{\'e}nic solitons coexist in the weakly relativistic regime, the ultrarelativistic EP plasmas in contrast support only the sub-Alfv{\'e}nic solitons. Different limits of the Mach numbers and soliton amplitudes are also examined in these two physical regimes. 
\end{abstract}
\begin{keyword}
Electron-positron plasma \sep Relativistic plasma \sep Alfv{\'e}nic  soliton \sep pseudopotential
\end{keyword}
\end{frontmatter}
\section{Introduction} \label{sec-intro}
Electron-positron (EP) plasmas have been known to play important roles in many physical situations,
 such as active galactic nuclei \citep{begelman1984, miler1987}, pulsars \citep{goldreich1969}, quasars \citep{wardle1998}, black holes \citep{blandford1977}, accretion disks \citep{orosz1997}, the early universe \citep{misner1973, gibbons1983}, near the polar cap of fast rotating neutron stars \citep{lightman1982, burns1982, lightman1987, yu1986}, as well as in laboratories \citep{sarri2015}. In the latter, it has been shown that the production of the ion-free high-density neutral EP-pair plasmas and their identification as collective modes can be possible in a controlled laboratory environment.    
 \par
  Linear and nonlinear waves in EP-pair plasmas differ fundamentally from those in ordinary electron-ion plasmas or from a purely electronic beam due to their intrinsic and complete symmetry with equal charge (but opposite in sign) and mass.   
The Sagdeev or pseudopotential approach has been  the most suitable technique for the desription of  nonlinear large amplitude waves  \citep{misra2011, misra2013, sagdeev1966, sagdeev1969, baboolal1990, mace1991, banerjee2015, banerjee2016, saini2011, das2010} which also works well  in  pair plasmas \citep{verheest1996}. However, when relativistic dynamics is included together with thermal pressure of plasma particles for the description of large amplitude EM waves, the Sagdeev's approach may not be suitable. In this context, an  alternative procedure has also been developed by McKenzie \textit{et al.} \citep{mckenzie2003} to study the properties of nonlinear waves in its own frame of reference.   Although both approaches are analogous to each other especially for electrostatic waves,  the McKenzie approach provides a better perception and usefulness than the Sagdeev's approach especially when one is concerned with the propagation  of electromagnetic (EM) solitary waves in plasmas \citep{verheest2004, verheest2005}.   In the latter,  Verheest and Cattaert have studied the propagation of large amplitude EM waves in nonrelativistic and relativistic EP-pair plasmas without any thermal flow  of electrons and positrons using the McKenzie approach.
\par
 In this work, our aim  is to advance and generalize the theory of Verheest and Cattaert \citep{verheest2004}   by considering the fully relativistic fluid models for electrons and positrons  which account  for physical regimes of both weakly relativistic    and ultrarelativistic   random thermal energies.
We show that in contrast to  the weakly realtivistic plasmas which support both sub-Alfv{\'e}nic and super-Alfv{\'e}nic solitons, only the sub-Alfv{\'e}nic  solitons can be formed in   EP-pair plasmas with ultrarelativistic energies.    
\section{Relativistic fluid model and multispecies integrals} \label{sec-fluid-model}
We consider the nonlinear propagation of EM solitary waves along the constant magnetic field $B_0\hat{x}$ in an EP-pair plasma with relativistic flow of thermal electrons and positrons. We assume that the effective collision frequency in an EP-pair plasma, which includes the recombination  and photon annihilation  effects, is assumed to be much smaller than the plasma oscillation frequency of electrons and positrons. 
 From the energy momentum tensor, the basic equations for the relativistic dynamics of a $j$-th species particle can be written as \citep{gratton1997,gomberoff1997} 
\begin{equation} \label{eq:basic1}
\dfrac{\partial}{\partial t}\left(\gamma_{j}n_{j}\right) +\nabla\cdot\left( \gamma_{j}n_{j}\textbf{v$_{j}$}\right)  =0,
\end{equation}
\begin{equation} \label{eq:basic2}
\begin{split}
\dfrac{H_{j}}{c^{2}}\left(\dfrac{\partial}{\partial t}+\textbf{v$_{j}$}\cdot\nabla\right)\left( \gamma_{j}\textbf{v$_{j}$}\right) &=n_{j}q_{j}\left( \textbf{E}+\dfrac{1}{c}\textbf{v$_{j}$}\times\textbf{B}\right)\\ &-\dfrac{1}{\gamma_{j}}\nabla P_{j}-\dfrac{\textbf{v$_{j}$}\gamma_{j}}{c^{2}}\dfrac{dP_{j}}{dt},
\end{split}
\end{equation}
\begin{equation} \label{eq:basic3}
\dfrac{1}{n_{j}}\dfrac{dP_{j}}{dt}=\dfrac{d}{dt}\left( \dfrac{H_{j}}{n_{j}}\right), 
\end{equation}
where $d/dt\equiv \partial_t+{\bf v}_j\cdot\nabla$,  $n_{j}$, $q_{j}$, $m_{j}$, ${\bf v}_j$, $\gamma_{j}$, $P_{j}$ and $H_{j}$ are, respectively, the number density, charge, mass, fluid velocity, relativistic factor, thermal pressure and   enthalpy per unit volume of $j$-species particle. Also,   ${\bf E}$ and ${\bf B}$ are the electric and  magnetic (total) fields respectively. Introducing ${\cal E}_j$ as  the total energy density and $\epsilon_j$ the internal energy density of the $j$-species fluid, we have    $H_{j}={\cal E}_j+P_{j}$  and ${\cal E}_j=n_{j}m_{j}c^{2}+\epsilon_{j}$.
We consider the ploytropic pressure law as  \citep{gratton1997,gomberoff1997}  $P_j=(\Gamma-1)\epsilon_j=n_jk_BT_j$, where $k_B$ is the Boltzmann constant, so that $\epsilon_j=n_jk_BT_j/(\Gamma-1)$ and $H_j\equiv n_j\alpha_j=n_jm_jc^2+\Gamma P_j/(\Gamma-1)=n_jm_jc^2[1+\Gamma\beta_j/(\Gamma-1)]$ with the energy ratio $\beta_j=k_BT_j/m_jc^2$ and the polytropic index  $4/3\leq\Gamma\leq5/3$.  In particular,  $\Gamma=5/3$ and $4/3$, respectively,  correspond  to the weakly relativistic (classical) and ultrarelativistic  regimes. 
So, in the  weakly relativistic  limit $P_j\ll n_jm_jc^2$ (applicable for low-energy plasmas)    we have for $\Gamma=5/3$, $H_j=n_jm_jc^2+(5/2)n_jk_BT_j\approx n_jm_jc^2$, and in the regime of ultrarelativistic energies where $P_j\gg n_jm_jc^2$, we have  instead $H_{j}=n_jm_jc^2+4n_jk_BT_j\approx 4n_jk_BT_j$.  
\par  
The system is then closed by the following Maxwell’s equations.

\begin{equation} \label{eq:max1}
\nabla\cdot\textbf{E}=4\pi\sum_{j}q_{j}n_{j}\gamma_{j},
\end{equation}
\begin{equation} \label{eq:max2}
\nabla\cdot\textbf{B}=0,
\end{equation}
\begin{equation} \label{eq:max3}
\nabla\times\textbf{E}=-\dfrac{1}{c}\dfrac{\partial\textbf{B}}{\partial t},
\end{equation}
\begin{equation} \label{eq:max4}
\nabla\times\textbf{B}=\dfrac{4\pi}{c}\sum_{j}q_{j}n_{j}\gamma_{j}\textbf{v$_{j}$}+\dfrac{1}{c}\dfrac{\partial\textbf{E}}{\partial t}.
\end{equation}
\par 
In order to derive an evolution equation for  purely stationary nonlinear  solitary EM waves and their properties from Eqs. \eqref{eq:basic1} to \eqref{eq:max4} we follow the McKenzie approach  as used in, e.g., Ref. \citep{verheest2004,verheest2005}. 
 First, we derive various conserved quantities for a general species $j$ before we apply it for an EP plasma. We look for the excitation of solitary waves that propagate along the constant magnetic field ${\bf B}_0$, i.e., the $x$-axis. In a frame moving with the constant speed $V$ along the $x$-axis, all plasma species have    the same constant velocity $V$ along the direction. Since in the wave frame there is no time derivative, 
   Eqs. (\ref{eq:basic1}) and (\ref{eq:basic2}) reduce to

\begin{equation} \label{eq:basic1_1}
\dfrac{d}{dx}\left( \gamma_{j}n_{j}\textbf{v$_{jx}$}\right)=0,
\end{equation}
\begin{equation} \label{eq:basic2_1}
\begin{split}
\dfrac{\alpha_{j}}{c^{2}}\gamma_{j}n_{j}v_{jx}\dfrac{d}{dx}\left( \gamma_{j}\textbf{v$_{j}$}\right) &=\gamma_{j}n_{j}q_{j}\left( \textbf{E}+\dfrac{1}{c}\textbf{v$_{j}$}\times\textbf{B}\right)\\& -\dfrac{dP_{j}}{dx}\hat{x}.
\end{split}
\end{equation}
Also, from Eqs. (\ref{eq:max1}) to (\ref{eq:max4}) we successively obtain the following equations.
\begin{equation} \label{eq:max1_1}
\dfrac{dE_{x}}{dx}=4\pi\sum_{j}q_{j}n_{j}\gamma_{j},
\end{equation}
\begin{equation} \label{eq:max2_1}
\dfrac{dB_{x}}{dx}=0,
\end{equation}
\begin{equation} \label{eq:max3_1}
\hat{x}\times\dfrac{d\textbf{E}}{dx}=0,
\end{equation}
\begin{equation} \label{eq:max4_1}
\hat{x}\times\dfrac{d\textbf{B}}{dx}=\dfrac{4\pi}{c}\sum_{j}q_{j}n_{j}\gamma_{j}\textbf{v$_{j}$}
\end{equation}
Now,  Eq. (\ref{eq:max2_1}) gives on integration $B_{x}=B_{0}$, a constant. Also,   from  Eq. (\ref{eq:max3_1}) it follows that $ {\bf E}_{\perp}=0$ under the boundary condition  ${\bf E}\rightarrow0$ as $x\rightarrow\pm\infty$, and so only $E_{x}= {d\phi}/{dx}$ ($\phi$ is the scalar potential) and ${\bf B}_\perp$ are the  variables,  which also tend to zero  as $x\rightarrow\pm\infty$, i.e., in the undisturbed plasma far away from the region of the nonlinear structure.   
Next, from the equation of  continuity  (\ref{eq:basic1_1}), we obtain the following conservation of  mass  (parallel flux).
\begin{equation} \label{eq:flux}
\gamma_{j}n_{j}v_{jx}=\gamma_{j0}n_{j0}V
\end{equation}
From Eq. (\ref{eq:basic2_1}), after summing over all the species and using Eqs. (\ref{eq:max1_1}),  (\ref{eq:max4_1}) and   (\ref{eq:flux}), we obtain
\begin{equation} \label{eq:basic2_2}
\begin{split}
\frac{V}{c^{2}}\sum_{j}\gamma_{j0}n_{j0}\alpha_{j}\frac{d}{dx}\left( \gamma_{j}\textbf{v$_{j}$}\right)&=\frac{1}{4\pi}\left[ E_{x}\frac{dE_{x}}{dx}+\left(\hat{x}\times\frac{d\textbf{B$_{\perp}$}}{dx}\right)\right.\\ 
&\left.\times\textbf{B}\right]-\sum_{j}\frac{dP_{j}}{dx}\hat{x}.
\end{split}
\end{equation}
  Integrating Eq. \eqref{eq:basic2_2} with respect to $x$ we obtain the following two distinct integrals of motion.
\begin{equation} \label{eq:IntegralOfMotion_1}
\begin{split}
\dfrac{V}{c^{2}}\sum_{j}\gamma_{j0}n_{j0}\alpha_{j}\left( \gamma_{j}{v_{jx}}-\gamma_{j0}V\right)& =\dfrac{1}{8\pi}\left(E_{x}^{2}-B_{\perp}^{2}\right) \\
&-\sum_{j}\left( P_{j}-P_{j0}\right), 
\end{split}
\end{equation}
\begin{equation} \label{eq:IntegralOfMotion_2}
\dfrac{V}{c^{2}}\sum_{j}\gamma_{j0}n_{j0}\alpha_{j}\gamma_{j}\textbf{v$_{j\perp}$} =\dfrac{B_{0}}{4\pi}\textbf{B$_{\perp}$}. 
\end{equation}
Furthermore, the projection of Eq. (\ref{eq:basic2_1}) on $ {\bf v}_{j\perp}$  gives
\begin{equation} \label{eq:projection_1}
\textbf{v$_{j\perp}$}\cdot\left[\dfrac{\gamma_{j}n_{j}\alpha_{j}}{c^{2}}\dfrac{d}{dx}\left( \gamma_{j}\textbf{v$_{j\perp}$}\right) \right] =\dfrac{q_{j}\gamma_{j}n_{j}}{c}\textbf{v$_{j\perp}$}\cdot\left( \textbf{e$_{x}$}\times\textbf{B$_{\perp}$}\right).
\end{equation}
Multiplying Eq. \eqref{eq:projection_1} by $\gamma_{j0} \alpha_j n_{j0}/q_{j}$,   summing over all the species and integrating we obtain
\begin{equation} \label{eq:projection_2}
\sum_{j}\dfrac{\gamma_j^2 \alpha_j^2 \gamma_{j0} n_{j0}}{q_{j}}v_{j\perp}^{2}=0,
\end{equation}
where we have used Eq. (\ref{eq:IntegralOfMotion_2}). 
We can also project Eq. (\ref{eq:basic2_1}) on $ {\bf v}_{j}$ to yield
\begin{equation} \label{eq:projection_3}
\dfrac{\alpha_{j}}{2c^{2}}\dfrac{d}{dx}\left( \gamma_{j}^{2}v_{j}^{2}\right) =q_{j}\gamma_{j}E_{x}-K_{B}T_{j}\dfrac{d}{dx}\left[ \log\left(  n_{j}\right) \right]. 
\end{equation}
\section{Relativistic EP plasmas: Energy integral} \label{sec-enrg-int}
We focus our attention to  an EP-pair plasma. The  results obtained in Sec. \ref{sec-fluid-model} will be modified with $q_{e}=-e$, $q_{p}=e$, $m_{e}=m_{p}=m$, $n_{e0}=n_{p0}=n_{0}$, $T_{e}=T_{p}=T$, $\gamma_{e0}=\gamma_{p0}=\gamma_{0}$ and $\alpha_{e}=\alpha_{p}=\alpha$, $\beta_e=\beta_p=\beta$, where the subscripts $j = e$ and $p$, respectively, stand for   electrons and   positrons.
Thus, for EP plasmas  the invariants (\ref{eq:flux}), (\ref{eq:IntegralOfMotion_1}), (\ref{eq:IntegralOfMotion_2}) and (\ref{eq:projection_2}), respectively,  reduce to
\begin{equation} \label{eq:flux_2}
\gamma_{e}n_{e}v_{ex}=\gamma_{p}n_{p}v_{px}=V\gamma_{0}n_{0},
\end{equation}
\begin{equation} \label{eq:IntegralOfMotion_1_1}
\begin{split}
\dfrac{V}{c^{2}}\gamma_{0}n_{0}\alpha\left( \gamma_{e}v_{ex}+\gamma_{p}v_{px}-2\gamma_{0}V\right) &=\dfrac{1}{8\pi}(E_x^2-B_{\perp}^{2})\\
& -\left( P_{e}+P_{p}-2P_{0}\right), 
\end{split}
\end{equation}
\begin{equation} \label{eq:IntegralOfMotion_2_1}
\dfrac{V}{c^{2}}\gamma_{0}n_{0}\alpha\left( \gamma_{e} {\bf v}_{e\perp}+\gamma_{p} {\bf v}_{p\perp}\right)  =\frac{B_{0}}{4\pi}{\bf B}_{\perp},  
\end{equation}
\begin{equation} \label{eq:projection_2_1}
\gamma_e^2 v_{e\perp}^{2}=\gamma_p^2 v_{p\perp}^{2}.
\end{equation}
Using Eq. (\ref{eq:projection_2_1}), we obtain from Eq. (\ref{eq:IntegralOfMotion_2_1}) the following two results
\begin{equation} \label{B_dot}
\left( \gamma_p \textbf{v$_{p\perp}$}-\gamma_e \textbf{v$_{e\perp}$}\right)\cdot\textbf{B$_{\perp}$}=0
\end{equation}
\begin{equation} \label{B_cross}
\left(\gamma_p \textbf{v$_{p\perp}$}+\gamma_e \textbf{v$_{e\perp}$}\right)\times\textbf{B$_{\perp}$}=0.
\end{equation}
Thus, it follows from Eqs.  (\ref{B_dot}) and (\ref{B_cross})   that while the component of $\left( \gamma_p \textbf{v$_{p\perp}$}-\gamma_e \textbf{v$_{e\perp}$}\right)$ is orthogonal to $ {\bf B}_\perp$,  the other component of $\left(\gamma_p \textbf{v$_{i\perp}$}+\gamma_e \textbf{v$_{e\perp}$}\right)$ is parallel to $ {\bf B}_\perp$.
\par In the weakly nonlinear theory, the truly stationary solutions are only possible at linear polarization of EM fields. So, we can assume  without loss of generality  that ${\bf B}_\perp$ is along the $y$-axis. Then Eqs. (\ref{B_dot}) and (\ref{B_cross}) give $\gamma_e v_{ey}=\gamma_p v_{py}$ and $\gamma_p v_{pz}=-\gamma_e v_{ez}$, and so the $y$-component of Eq. (\ref{eq:max4_1}) gives 
\begin{equation}
\sum_{j=e,p} q_jn_j\gamma_j v_{jy}=0, \label{eq-quasin0}
\end{equation}
 from which one obtains $n_e=n_p=n$, say. So, form Eq. (\ref{eq:flux_2}) we have $\gamma_e v_{ex}=\gamma_p v_{px}$, and  using  the charge neutrality condition, $\gamma_e n_e=\gamma_p n_p$  we have $\gamma_e=\gamma_p=\gamma$, say. Thus, we have $v_{ex}=v_{px}=v_x$, $v_{ey}=v_{py}=v_y$, $v_{ez}=-v_{pz}=v_z$ and $v_p^2=v_e^2=v^2$. 
Using the charge neutrality condition, the  Amp{\'e}re law (\ref{eq:max4_1}) reduces to
\begin{equation}
\textbf{e$_{x}$}\times\dfrac{d \textbf{B$_{\perp}$}}{dx}=\dfrac{4\pi e n\gamma}{c}\left( \textbf{v$_{i\perp}$}-\textbf{v$_{e\perp}$}\right),
\end{equation}
where $\gamma=1/\sqrt{1-v^2/c^2}$. 
Furthermore, a scalar multiplication of ${d \textbf{B$_{\perp}$}}/{dx}$ with $\textbf{B$_{\perp}$}$ gives
\begin{equation}
\textbf{B$_{\perp}$}\times\frac{d \textbf{B$_{\perp}$}}{dx}=0,
\end{equation}
meaning that the wave magnetic field $\textbf{B$_{\perp}$}$ is linearly polarized. Thus, our assumption of linear polarization of EM fields and quasineutrality condition are valid.   Next, from Eqs. 
(\ref{eq:IntegralOfMotion_1_1}) and (\ref{eq:IntegralOfMotion_2_1}), we obtain
\begin{equation} \label{eq:29}
\begin{split}
\gamma v_{x}&=\dfrac{1}{2}\left[ \gamma_{0}V-\dfrac{mc^{2}V_{A}^{2}}{2V\gamma_{0}\alpha}b^{2}+\dfrac{c^{2}K_{B}T}{V\gamma_{0}\alpha}\right.\\
&\left.+\sqrt{\left( \gamma_{0}V-\dfrac{mc^{2}V_{A}^{2}}{2V\gamma_{0}\alpha}b^{2}+\dfrac{c^{2}K_{B}T}{V\gamma_{0}\alpha}\right) ^{2}-\dfrac{4c^{2}K_{B}T}{\alpha}}\right], 
\end{split}
\end{equation}
\begin{equation} \label{eq:30}
\gamma v_{y}=\dfrac{mc^{2}V_{A}^{2}}{V\gamma_{0}\alpha}b
\end{equation}
where $b=B_{y}/B_{0}$ is the dimensionless wave magnetic field, $\gamma_0=1/\sqrt{1-V^2/c^2}$ and   $V_{A}$ is the Alfv\'en   velocity in an EP plasma, defined by, $V_{A}^{2}=B_{0}^{2}/(8\pi n_{0}m)$.
Next, rearranging the $y$-component of Eq. (\ref{eq:basic2_1}), we obtain another velocity component 
\begin{equation} \label{eq:31}
v_{z}=\dfrac{\alpha v_{x}}{ceB_{0}}\dfrac{d}{dx}\left( \gamma v_{y}\right) 
\end{equation}
Note that  Eq. (\ref{eq:flux_2}) results into $n=(V\gamma_{0}n_{0})/(\gamma v_{x})$ which when applied to  Eq. (\ref{eq:projection_3}) gives, after integration and   summation over electron and positron species, the following conservation of kinetic energy.
\begin{equation} \label{eq:32}
\gamma^{2}\left(v_{x}^{2}+v_{y}^{2}+v_{z}^{2}\right)=\gamma_{0}^{2}V^{2}-\dfrac{2c^{2}K_{B}T}{\alpha}\log\left( \dfrac{\gamma_{0}V}{\gamma v_{x}}\right). 
\end{equation}
 We define the  Mach number as $M=V/V_{A}$ and a dimensionless coordinate $\zeta=x\omega_{p}/c$, where $\omega_{p}^{2}\equiv \omega_{pe}^2+\omega_{pp}^2=8\pi n_{0}e^{2}/m$ is the squared  total  plasma oscillation frequency  of electrons and positrons. Finally,   using Eqs. (\ref{eq:29}) to (\ref{eq:31}), we obtain from  Eq. (\ref{eq:32})  the following equation.
\begin{equation} \label{eq:EnergyIntegral}
\dfrac{1}{2}\left( \dfrac{db}{d\zeta}\right)^{2}+\psi(b) =0,
\end{equation}
where $\psi$ is the Sagdeev potential or pseudopotential, given by,
\begin{equation} \label{pseudopotential} 
\psi(b)=\dfrac{M^{2}}{2}\left( 1-\dfrac{f}{g^{2}}\right)
\end{equation}  and
 \begin{equation} \label{eq:f1}
f=1-\dfrac{m^{2}c^{4}}{\gamma_{0}^{4} M^{4}\alpha^{2}}b^{2}+\dfrac{2c^{2}K_{B}T}{\alpha\gamma_{0}^{2}V^{2}}\log\left( g\right),
\end{equation}
\begin{equation} \label{eq:xi1}
\begin{split}
g&=\dfrac{1}{2}\left[ 1-\dfrac{mc^{2}}{2 \gamma_{0}^{2} M^{2}\alpha}b^{2}+\dfrac{c^{2}K_{B}T}{\alpha\gamma_{0}^{2}V^{2}}\right.\\
&\left.+\sqrt{\left( 1-\dfrac{mc^{2}}{2 \gamma_{0}^{2} M^{2}\alpha}b^{2}+\dfrac{c^{2}K_{B}T}{\alpha\gamma_{0}^{2}V^{2}}\right) ^{2}-\dfrac{4c^{2}K_{B}T}{\alpha\gamma_{0}^{2}V^{2}}}\right]. 
\end{split}
\end{equation}
Equation (\ref{eq:EnergyIntegral}) represents an energy integral for a pseudo particle of unit mass at pseudo time $\zeta$ moving with the pseudo velocity $db/d\zeta$ with a pseudopotential energy $\psi(b)$. 
In particular, in absence of the  effects of  relativistic flow $(\gamma_0\sim1)$ and thermal pressures of electrons and positrons $(\beta\sim0)$,       the pseudopotential  [Eq. \eqref{pseudopotential}]    reduces to
\begin{equation}
\psi(b)=\dfrac{M^{2}}{2}\left[ 1+\dfrac{4\left( b^{2}-M^{4}\right) }{\left(b^{2}-2M^{2}\right)^{2}}\right]
\end{equation}
which is exactly   the same as in Ref. \citep{verheest2004}. Introducing the parameter $v_{0}=V/c$ and noting that   $\beta \equiv K_{B}T/mc^{2}\ll 1$ defines the regimes of weakly relativistic (classical) plasmas and $\beta\gg 1$  that of ultra-relativistic plasmas,  we recast $f$ and $g$ as 
\begin{equation}
f=1-\dfrac{b^{2}(1-v_0^2)^2}{M^{4}  \left[1+\Gamma\beta/(\Gamma-1)\right] ^{2}}+2S \log g, \label{eq-f}
\end{equation}
\begin{equation}
\begin{split}
g&=\dfrac{1}{2}\left[ 1-\frac{b^{2}(1-v_0^2)}{2M^{2}  \left[1+\Gamma\beta/(\Gamma-1)\right]}+S+\right.\\
&\left.\sqrt{\left( 1-\dfrac{b^{2}(1-v_0^2)}{2M^{2}  \left[1+\Gamma\beta/(\Gamma-1)\right]}+S\right) ^{2}-4S}\right], \label{eq-g}
\end{split}
\end{equation} 
where $S={(1-v_{0}^{2})\beta}/{v_{0}^{2}\left[1+\Gamma\beta/(\Gamma-1)\right]}$.
\par 
A general discussion of Eq. \eqref{eq:EnergyIntegral} is almost similar to the Sagdeev's approach for large amplitude nonlinear waves.   The necessary conditions for the existence of solitary waves are (i)     $\psi(b)=0$   and $d\psi/db=0$ at $b=0$, (ii) $d^2\psi/db^2<0$ at $b=0$ (iii) $\psi(b_m\neq0)=0$, $\psi(b)<0$ for $0<|b|<|b_m|$ and $\left(d\psi/db\right)\vert_{b=b_m}\gtrless0$ according to when the solitary waves are compressive (with $b>0$) or rarefactive (with $b<0$). Here, $b_m$ corresponds to the amplitude of the solitary waves.   It is straightforward to verify that the condition  (i) is  satisfied. However, the condition (ii)   is satisfied for $M>M_c$, where $M_c$ is the critical value of $M$, given by,  
\begin{equation}
M_c=\sqrt{\frac{1-v_0^2}{1+ \Gamma\beta/(\Gamma-1)}}. \label{eq-Mc}
\end{equation}
Later, we will verify the condition (iii) numerically in two different regimes, i.e.,  weakly relativistic and ultrarelativic regimes. Furthermore, since the pseudopotential   $\psi(b)$ is to be a real valued function,  the expression under the square root in $g$ must be either zero or positive, yielding $\vert b \vert <|b_m|\leq b_c$ where  
\begin{equation}
b_c=\sqrt{2} M \left(1-\sqrt{S}\right)\sqrt{\frac{1+\Gamma\beta/(\Gamma-1)}{1-v_0^2}}.
\end{equation}
It follows that for some given values of  $M$, $\beta$ and $v_0$, the wave amplitude will not exceed the critical value $b_c$.  The upper limit of the Mach number $M_u$   can be obtained in terms of $\beta$ and $v_0$ from the condition  $\psi(b_c)\geq0$  as 
\begin{equation}
M_u=\frac{\sqrt{2(1-v_0^2)}\left(1-\sqrt{S}\right)}{\sqrt{\left[1+\Gamma\beta/(\Gamma-1)\right]\left[1+S (\log S-1)\right]}}. \label{eq-Mu}
\end{equation}
Thus, in order that the EP plasmas support large amplitude solitary waves, we must have    $M_c<M<M_u$.
In particular, for $\beta\rightarrow0$ (cold plasmas) and  $\gamma_0\sim1$, i.e., $v_0\ll1$ (nonrelativistic plasmas) we have $M_c\sim1$ and $M_u\sim\sqrt{2}$, i.e.,    super-Alfv{\'e}nic  solitons may exist with the Mach number satisfying $1<M<\sqrt{2}$.    This is in agreement with the results of Verheest and Cattaert   \citep{verheest2004}, who reported in nonrelativistic cold   electron-positron   plasmas. 
Next, in order that   $M_c<M_u$ holds, the function $A(v_0,\beta)$ must be positive, where
\begin{equation} \label{A}
A(v_0,\beta)=2(1-\sqrt{S})^2-1-S(\log S-1),  
\end{equation}
together with $0<S<1$.
In what follows, we examine numerically  the conditions and different limits of the wave amplitude  and the Mach number stated above for the existence of large amplitude EM solitons.   We  focus our discussion on two particular physical  regimes of weakly relativistic   $(\beta\ll1)$ and   ultrarelativistic  $(\beta\gg1)$ plasmas. These are demonstrated in the   two subsections \ref{sec-nonrelati} and \ref{sec-ultrarelati}.   Note that  one can, in principle, consider some other finite values of $\beta$, which may be neither much smaller nor  much larger than unity, however, a corresponding choice of the polytropic index in between $4/3\le\Gamma\le5/3$ may not be appropriate, and can lead to some incorrect results.   
\subsection{Weakly relativistic regime ($\beta \ll 1$)} \label{sec-nonrelati}
\begin{figure}[h!]
\centering
\includegraphics[width=3.5in,height=2in]{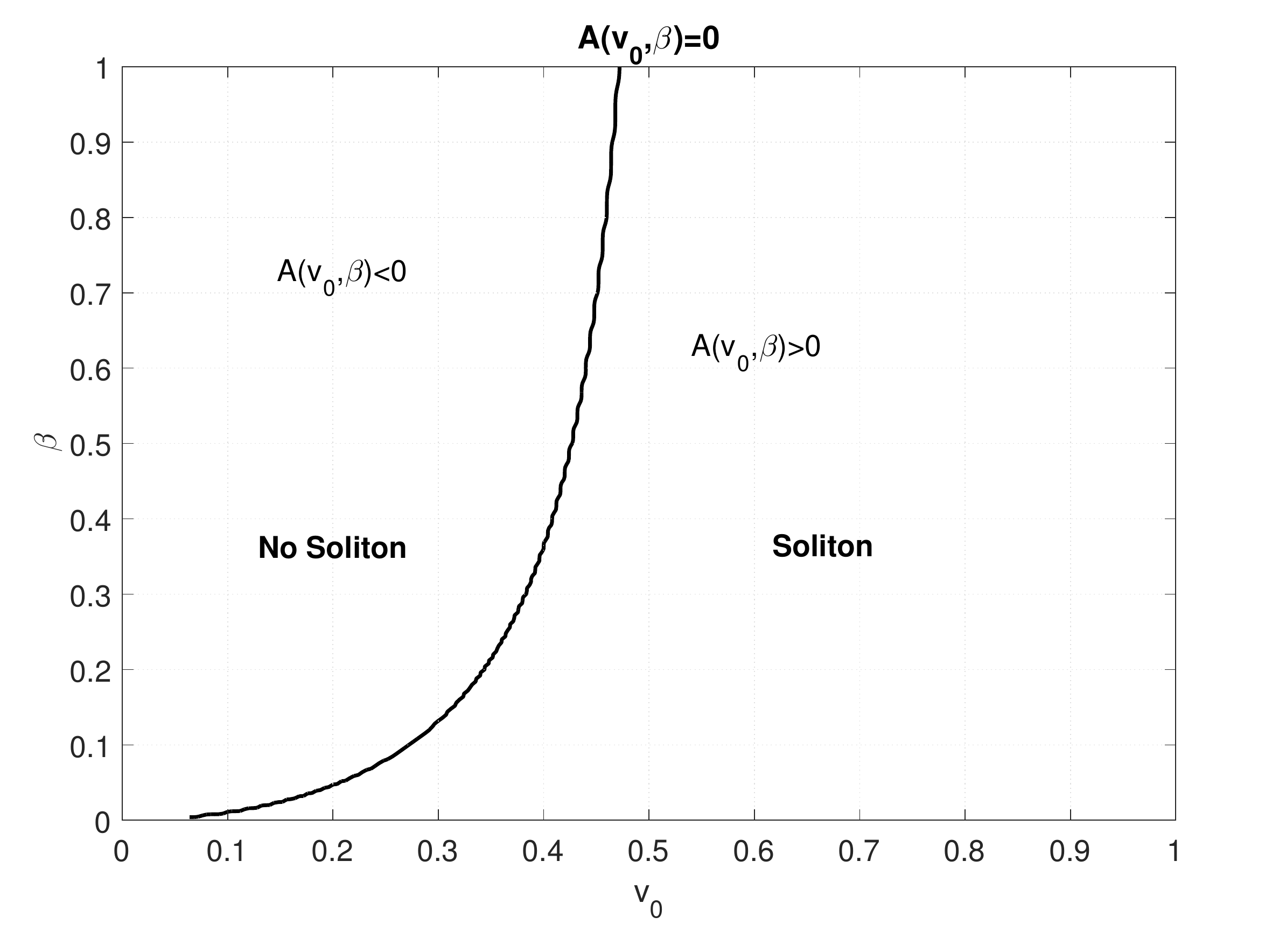}
\caption{$A(v_0,\beta)=0$ [Eq. \eqref{A}] is contour plotted to show the existence and non-existence regions of EM solitary waves in weakly relativistic $(\beta\ll1)$ plasmas.}
\label{fig:A_plot_cls}
\end{figure}

We consider $\Gamma=5/3$.  Since $0<S<1$ and $0<\beta \ll 1$, we have two cases of interest (i) $0<v_0<\sqrt{2/9}$, $0<\beta<v_0^{2}/(1-7v_0^{2}/2)$, i.e., when the upper limits of $\beta$ depend  on $v_0$     and (ii) $\sqrt{2/9}\leq v_0<1$, $0<\beta \ll 1$, i.e., when the upper limit of $\beta$ is independent of $v_0$.  Figure \ref{fig:A_plot_cls} is the contour plot of $A(v_0,\beta)=0$ showing the possible existence region of solitary waves in the $(v_0,\beta)$-plane. Within the domain $0<v_0<\sqrt{2/9}$, the ranges of values of $\beta$ change according to case (i). For example, the admissible range of $\beta$ at $v_0=0.3$ is $0<\beta<0.13$ and at $v_0=0.4$ it is $0<\beta<0.36$. So, smaller the values of $v_0$, lower is the  upper limit of  $\beta$.    On the other hand, when $\sqrt{2/9}\leq v_0<1$ and $\beta$ is independen on $v_0$,   there is a wide range  of values of $\beta:~0<\beta\ll1$  for which the solitary waves exist.   However, in all the domains the   solitary waves must have a maximum amplitude $b_c$, provided the admissible Mach number lies in $M_c<M<M_u$. 
\par
\begin{figure}[h!]
\includegraphics[scale=0.38]{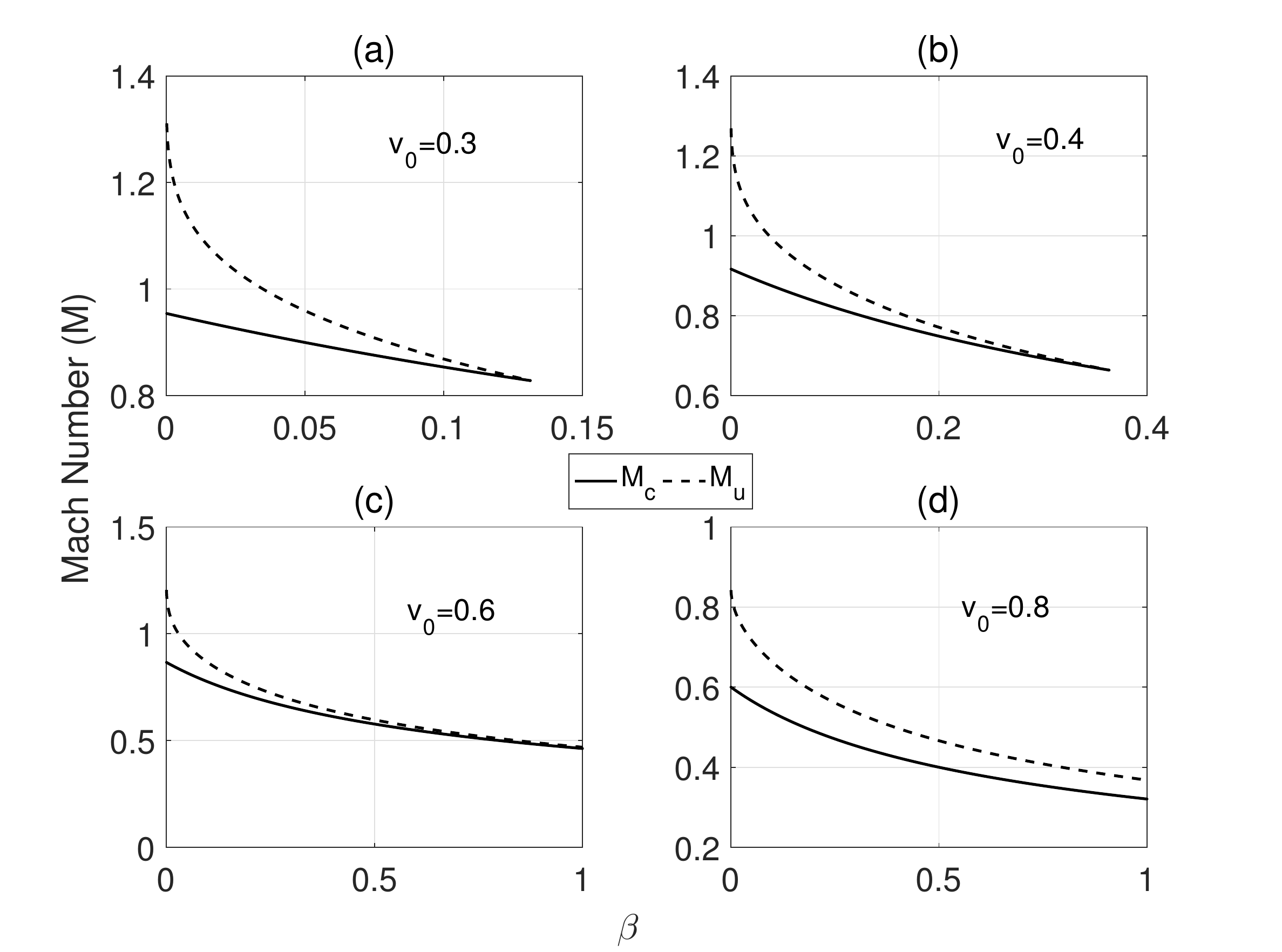}
\caption{Plots of the lower $(M_c)$ and upper $(M_u)$ limits of the Mach number, given by Eqs. \eqref{eq-Mc} and \eqref{eq-Mu}, are shown   for different values of $v_0$ in weakly relativistic $(0<\beta\ll1)$ plasmas . The subplots (a) and (b)  correspond  to the regimes   $0<v_0<\sqrt{2/9}$, $0<\beta<v_0^{2}/(1-7v_0^{2}/2)$, while the subplots (c) and (d) for  $\sqrt{2/9}\leq v_0<1$, $0<\beta \ll 1$. Note that $M_c,~M_u\leq 1$ for $0.7<v_0<1$. }
\label{fig:McMu_cls}
\end{figure}

 
Figure \ref{fig:McMu_cls} displays the plots of the lower (solid line) and upper (dashed line) limits of the Mach number   within the domain $0\leq\beta<1$  for different values of $v_0$ in two cases discussed before [\textit{cf}. Fig. \ref{fig:A_plot_cls}]. The subplots (a) and (b) correspond to the case (i) where $\beta$ depends on $v_0$, while (c) and (d) that for the case (ii) where $\beta$ does not depend on $v_0$.    We note that the values of $M_c$  are always less than unity, while those of $M_u$ can be less than or greater than unity depending on the values of $\beta$ and $v_0$ within the regimes. Here, the values of $\beta$ at which both $M_c$ and $M_u$ coincide are not admissible, because otherwise $M=M_c=M_u$ would violate  the condition for the existence of solitary waves.  If we  scale $\beta\lesssim0.05$ to interpret its smallness in the weakly relativistic regime, then from the subplots (a) and (b) of Fig. \ref{fig:McMu_cls}  we find that there are, in fact, two subregimes of $\beta$, namely $0<\beta<\beta_1$ and $\beta_1<\beta\lesssim0.05$. In the former regime, we have $1<M_u<1.4$, while in the other one has $0<M_u<1$. The threshold value $\beta_1$ shifts towards lower values  as the value of $v_0$ is increased within the admissible domain.  In fact, for values of $v_0\gtrsim0.7$, the threshold value disappears and only we have $0<M_u<1$ in $0<\beta\lesssim0.05$. 
 Thus, it follows that the   EP plasmas with weakly realativistic $(0<\beta\lesssim0.05)$ energies can support both the sub-Alfv{\'e}nic $(0<M<1)$ and super-Alfv{\'e}nic $(1<M<1.4)$ solitons in the regime $0<v_0<0.7$,  while only the sub-Alfv{\'e}nic solitons may exist for $0.7\lesssim v_0<1$. 
  From Fig. \ref{fig:McMu_cls},  it is also noticed that the values of both $M_c$ and $M_u$ decrease with increasing values of $v_0$,  and  they tend to become smaller than unity as $v_0$ approaches $1$, implying that as the phase velocity of EM solitary waves approaches the speed of light in vacuum, it is more likely that the sub-Alfv{\'e}nic solitons can exist in relativistic EP-pair plasmas. 
\par
\begin{figure}[h!]
\includegraphics[scale=0.36]{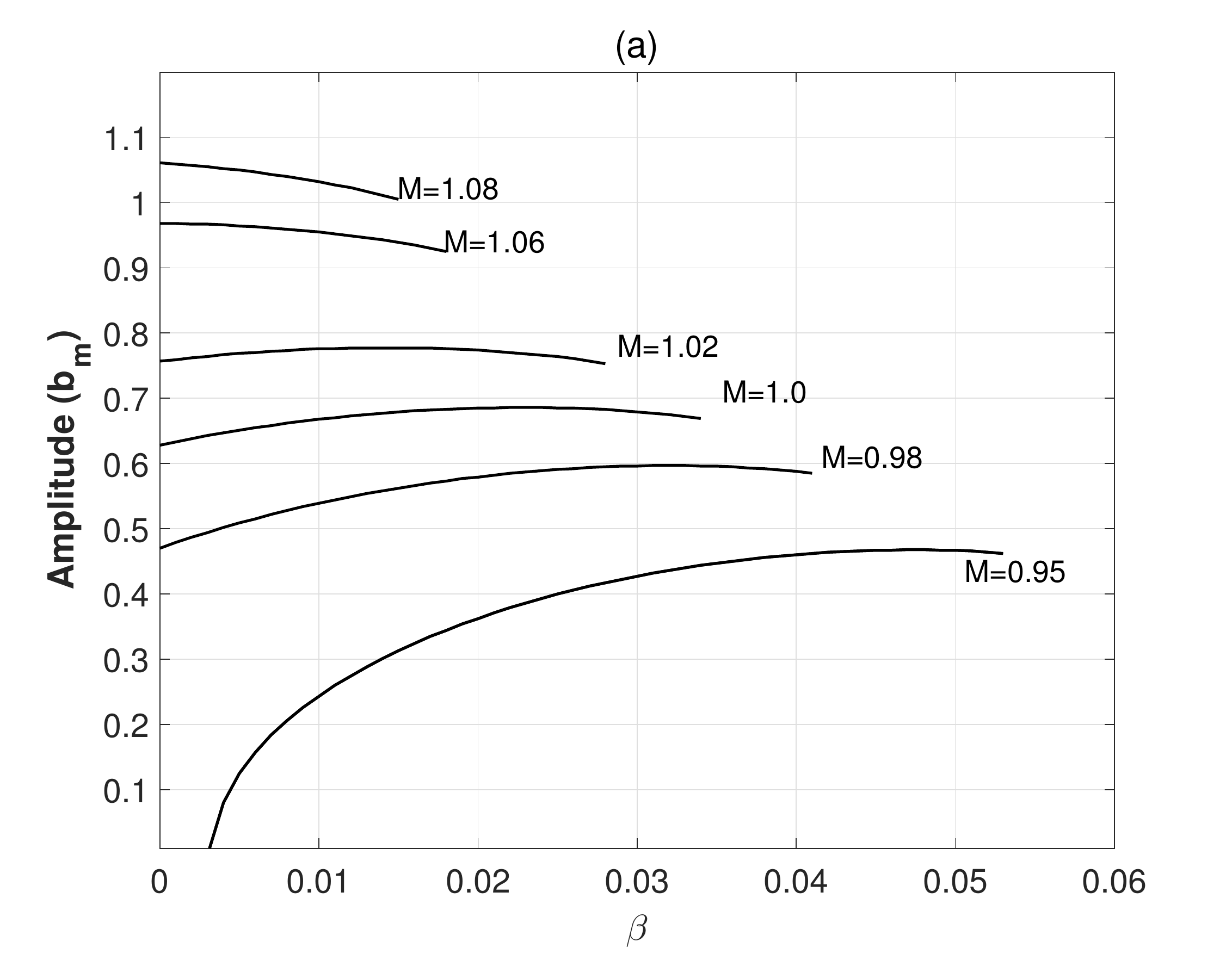}
\includegraphics[scale=0.36]{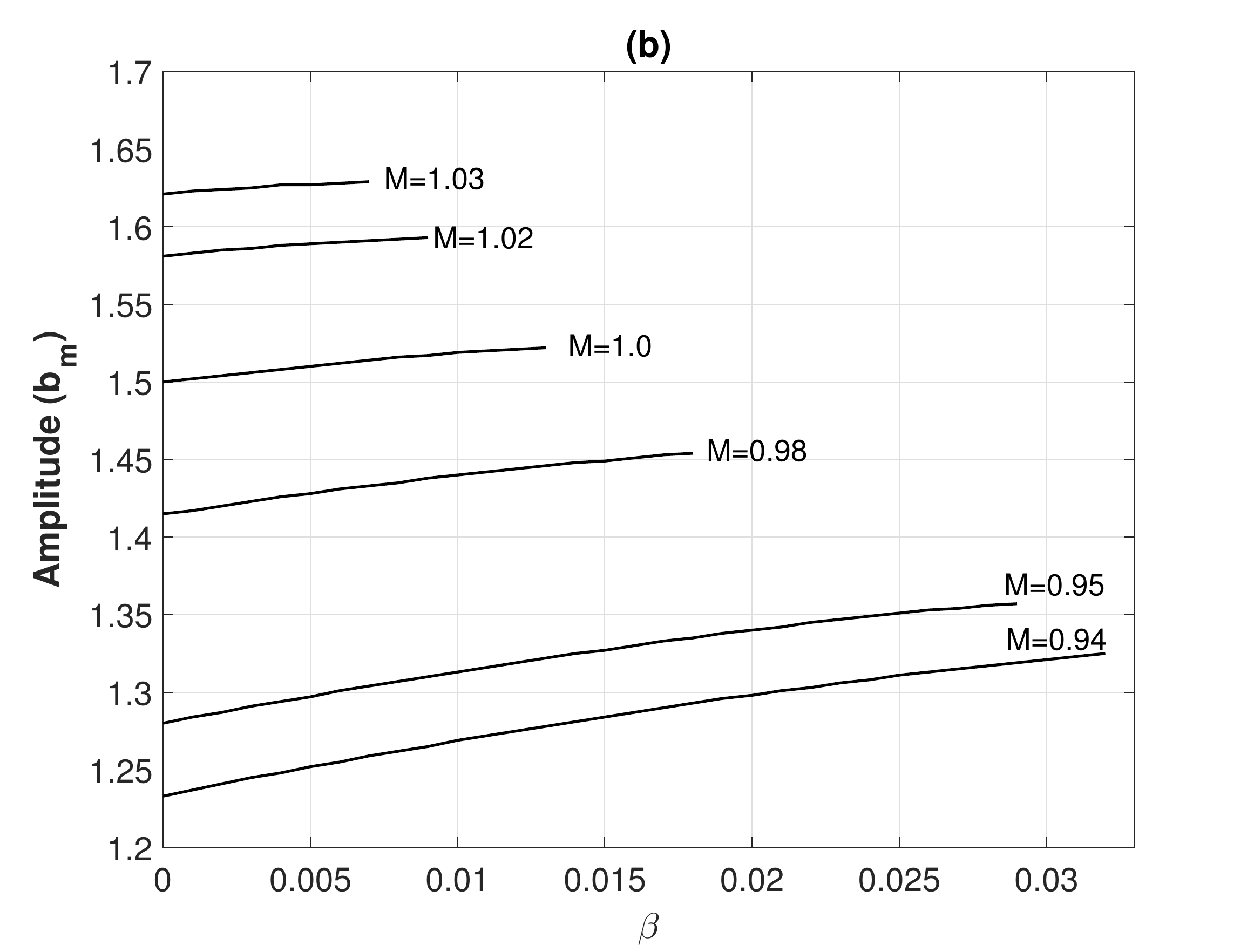}
\caption{The soliton amplitude $b_m$  is shown against $\beta$ with the variations of the Mach number $M$ in two different regimes of $v_0$ and $\beta$:   (a) $v_0=0.3$ within $0<v_0<\sqrt{2/9}$; $0<\beta<v_0^{2}/(1-7v_0^{2}/2)$, and (b) $v_0=0.6$ within  $\sqrt{2/9}\leq v_0<1$; $0<\beta \ll 1$. Note that the range   of  $\beta$ where $b_m$ is defined  differs and increases with decreasing values of $M$.  }
\label{fig:amp_beta_v03}
\end{figure}

 
In what follows, we numerically examine the variations of the wave amplitude $b_m$ [at which $\psi(b)=0$] against the parameter $\beta$ $(0\leq\beta\ll1)$  for different values of the Mach number, $M_c<M<M_u$ and with two different values of $v_0$, taking one from each of the regimes $0<v_0<\sqrt{2/9}$ and $\sqrt{2/9}<v_0<1$.   In these regimes of $M$ and $\beta$, the values of $b_m$ are always found to be $\lesssim b_c$.  We consider  (a) $v_0=0.3$ when the upper limit of $\beta$ depends on $v_0$, i.e., $0<\beta<v_0^{2}/(1-7v_0^{2}/2)$  and (b) $v_0=0.6$ when    $\beta$ does not depend on $v_0$. The results are shown in Fig.  \ref{fig:amp_beta_v03}. It mainly displays the contour plots of $\psi(b_m\neq0)=0$ in the $(\beta, b)$-plane.  It is interesting to note from subplot (a) that within the domain $0\lesssim\beta<0.06$ and  for a fixed   value of $v_0=0.3$  in $0<v_0<\sqrt{2/9}$,   the amplitude $b_m$ increases with increasing values of $M$ in $M_c<M<M_u$. However, the same decrease with increasing values of $\beta$  until  $1.06\lesssim M<M_u$. However,  as $M$  decreases from $M=1.06$ to lower values within the domain $M_c<M<1.06$,   the values of $b_m$ increase in a subinterval $0\leq\beta\lesssim\beta_2$, while those decrease in an other subinterval $\beta_2<\beta<0.06$.    Here, $\beta_2$ is some threshold value of $\beta$ which shifts to higher values as  $M$ decreases from $1.06$ to $M_c$.   On the other hand, for a fixed value $v_0=0.6$ in $\sqrt{2/9}<v_0<1$ [subplot (b)], the wave amplitude always increases with increasing values of both $\beta$   $(0\lesssim\beta<0.05)$ and  $M$   $(M_c<M<M_u)$. From the subplots (a) and (b) it is also seen that  the ranges of values of  $\beta$ where $b_m$ is defined  differ  and increase  with decreasing values of $M$.
\par
\begin{figure*}
\includegraphics[scale=0.34]{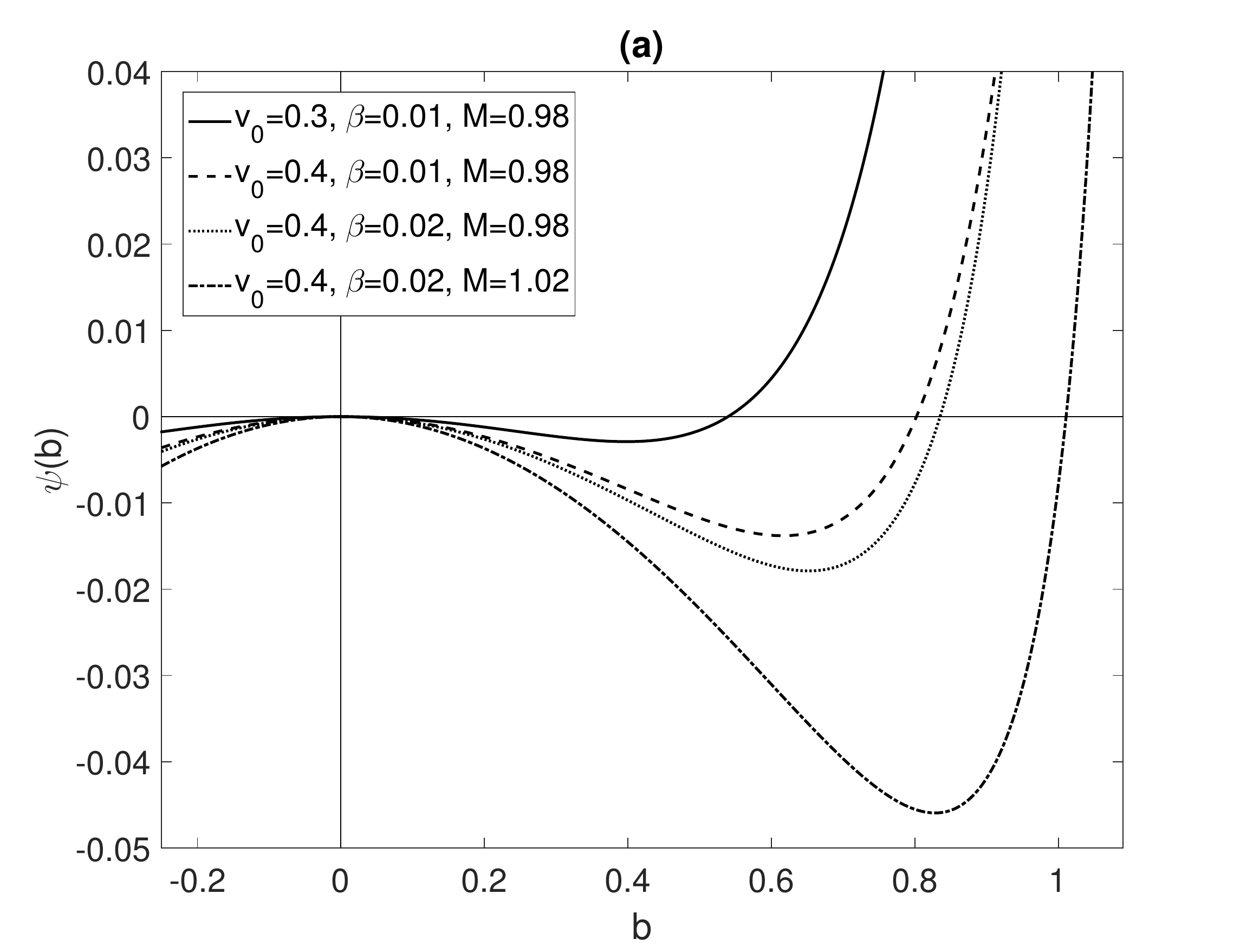}
\includegraphics[scale=0.34]{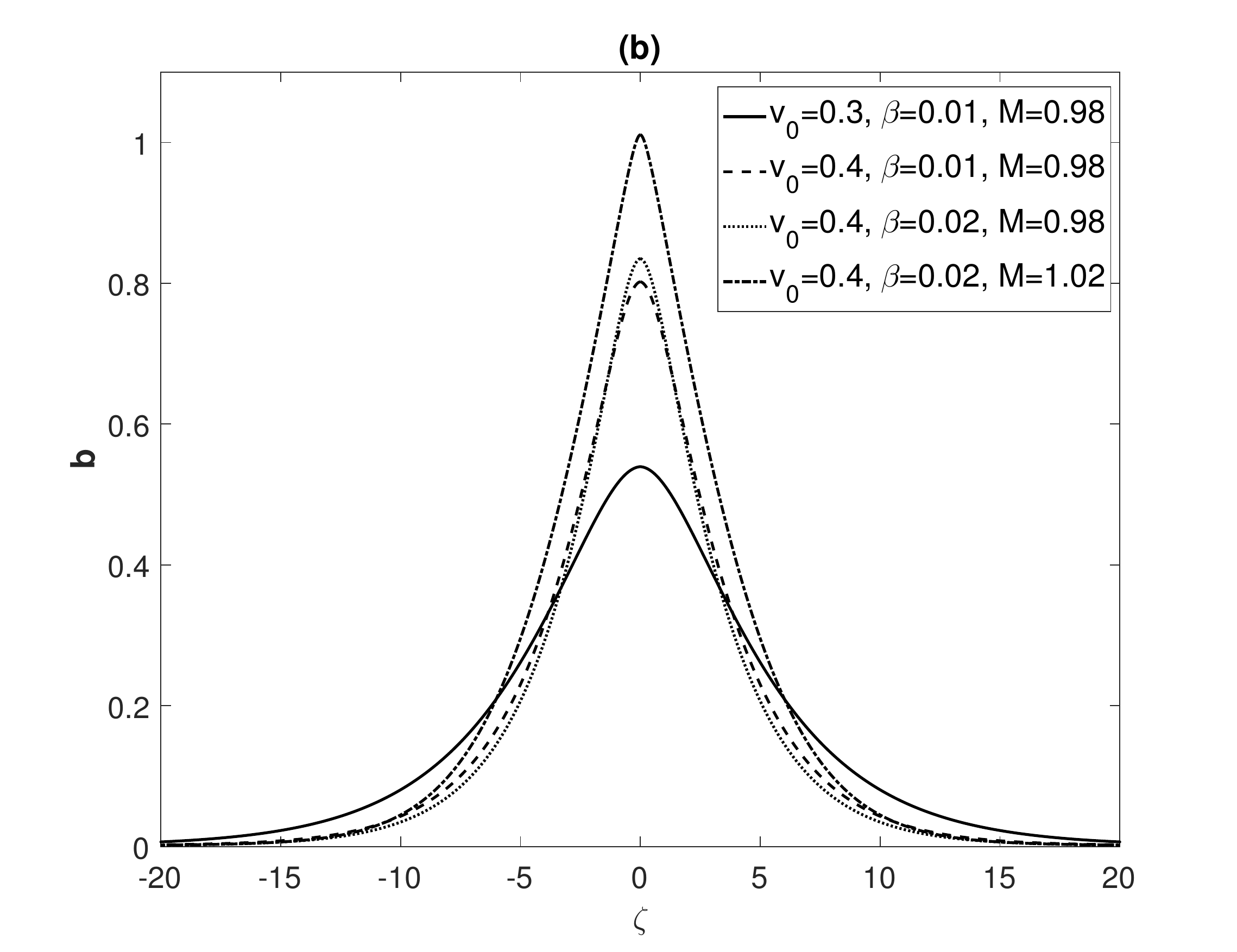}
\includegraphics[scale=0.34]{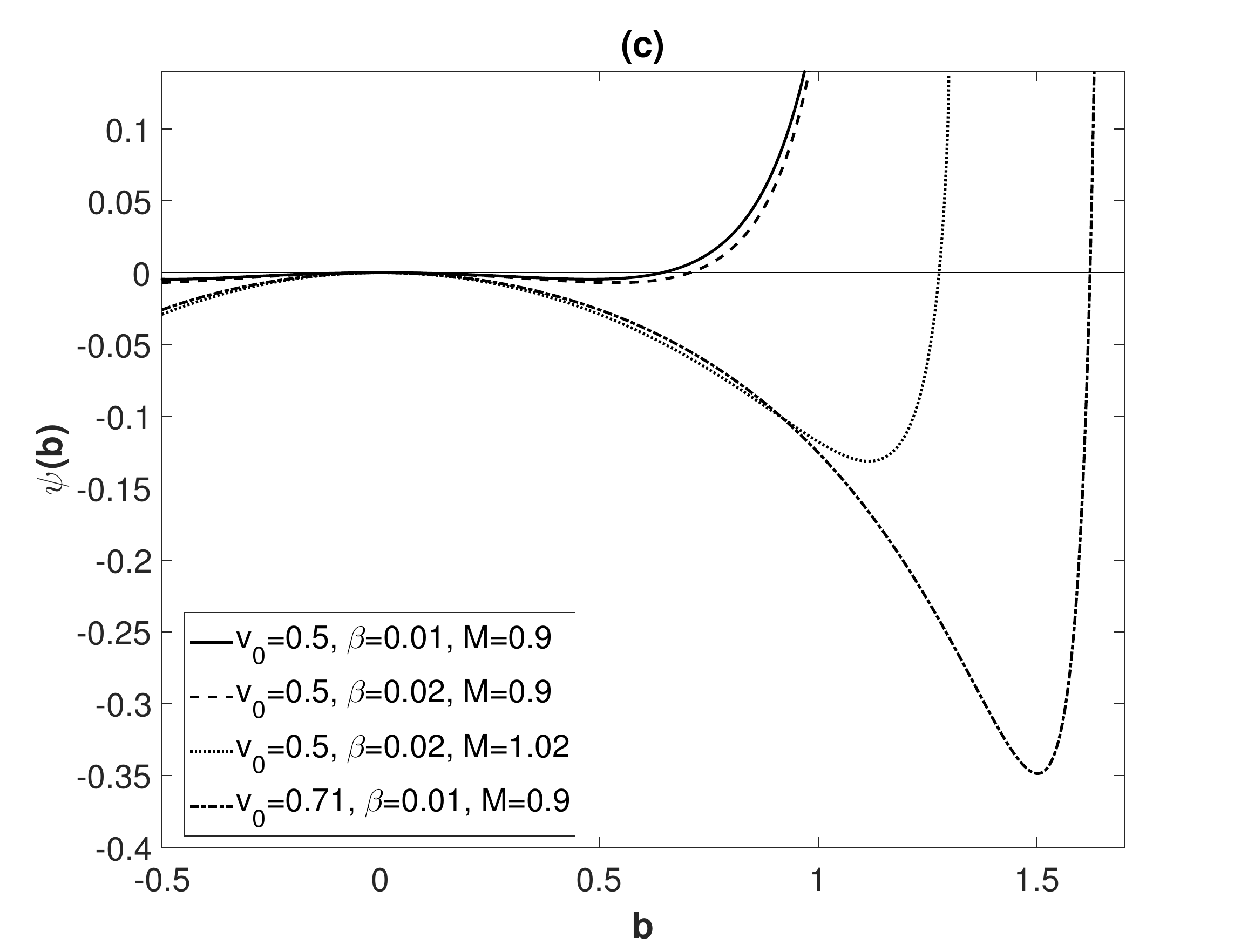}
\includegraphics[scale=0.34]{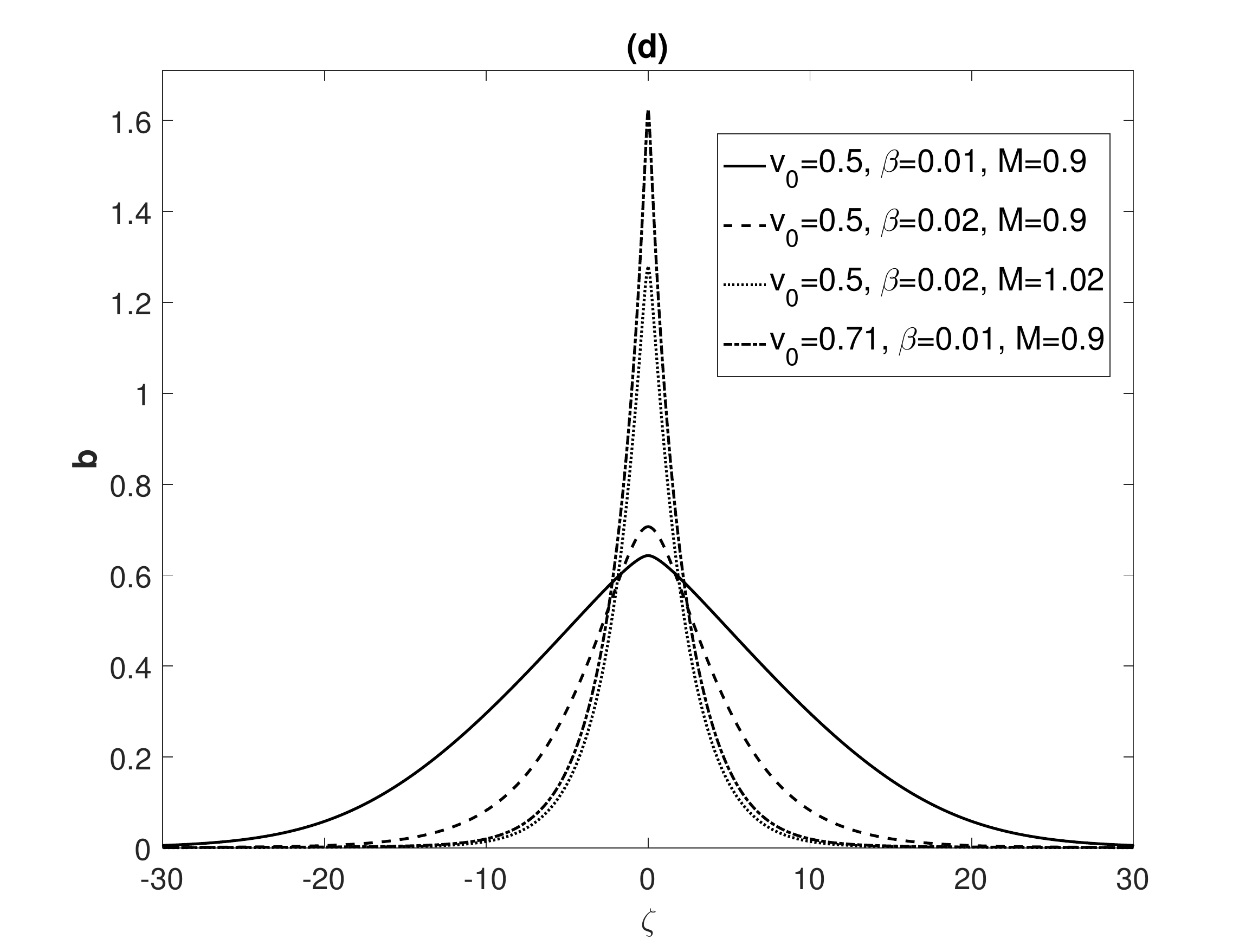}
\caption{Plots of the pseudopotential $\psi(b)$ [subplots (a) and (c)] and the corresponding soliton profile  [(b) and (d)] for different values of $v_0$, $\beta$ and $M$ as in the legends in two different regimes: (i) $0<v_0<\sqrt{2/9}$, $0<\beta<v_0^{2}/(1-7v_0^{2}/2)$, $M_c<M<M_u$ [subplots (a) and (b)] and (ii)    $\sqrt{2/9}\leq v_0<1$, $0<\beta \ll 1$, $M_c<M<M_u$ [subplots (c) and (d)].}
\label{fig:psi_cls_Mg1_v03}
\end{figure*}

 
Having obtained various parameter regimes for the existence of EM solitary waves as discussed before, we now plot the profiles of the pseudopotential $\psi(b)$ and the corresponding solitary structures as in  Fig. \ref{fig:psi_cls_Mg1_v03} for different values of $v_0$,  $\beta$ and the Mach number $M$ in two different regimes (i) $0<v_0<\sqrt{2/9}$, $0<\beta<v_0^{2}/(1-7v_0^{2}/2)$, $M_c<M<M_u$ [subplots (a) and (b)] and (ii)    $\sqrt{2/9}\leq v_0<1$, $0<\beta \ll 1$, $M_c<M<M_u$ [subplots (c) and (d)]. 
 As expected, the amplitudes of the solitons exactly correspond  to the cut-off values of $\psi$ at $b=b_m\neq0$ (i.e., the points where $\psi$ crosses the $b$-axis). From the profiles of $\psi$ and $b$,  the soliton widths   can also be verified by the formula:   width $W=|b_m/\psi_\text{min}|$.   An enhancement of  the amplitude and broadening of the  soliton profile (width) are seen to occur with an increase of the Mach number, however,  the amplitude increases but the width decreases with increasing value of $v_0$ and  $\beta$ within  the admissible regimes [subplots (a) and (b)]. On the other hand, subplots (c) and (d)  show   the same qualitative behaviors, i.e.,  with an increase of any one of $v_0$, $\beta$ and $M$,   the amplitude increases and the width decreases.       
\subsection{Ultra-relativistic regime ($\beta \gg 1$)} \label{sec-ultrarelati}
We consider the polytropic index $\Gamma=4/3$. In this case, since $0<S<1$ and $\beta \gg 1$, we can have also two possible regimes similar to the weakly relativistic case, namely (i) $\sqrt{1/6}<v_0<\sqrt{1/5}$, $1\ll\beta<v_0^{2}/(1-5v_0^{2})$, i.e., when the upper limits of $\beta$ depend  on the values of $v_0$     and (ii) $\sqrt{1/5}<v_0<1$, $\beta \gg 1$, i.e., when the upper limits of $\beta$  do  not depend  on $v_0$.    However, looking at the expressions of $M_c$ and $M_u$, we find that within the regime $\sqrt{1/6}<v_0<\sqrt{1/5}$, the ratio $M_u/M_c=\sqrt{2}(1-\sqrt{S})/\sqrt{1+S(\log S-1)}$ varies from $0.9814$ to $0.9996$, i.e.,  $M_u/M_c\sim1$ for $\beta\gg1$.  A numerical estimation also reveals that in this regime of $v_0$, $|\psi(b)|\lesssim10^{-9}$ and the soliton amplitude $|b_m|\lesssim0.01$. So, we are not interested in this short regime of $v_0$, and only the regime to be considered  for analysis is $\sqrt{1/5}<v_0<1$, $\beta \gg 1$.     
\begin{figure*}
\includegraphics[scale=0.36]{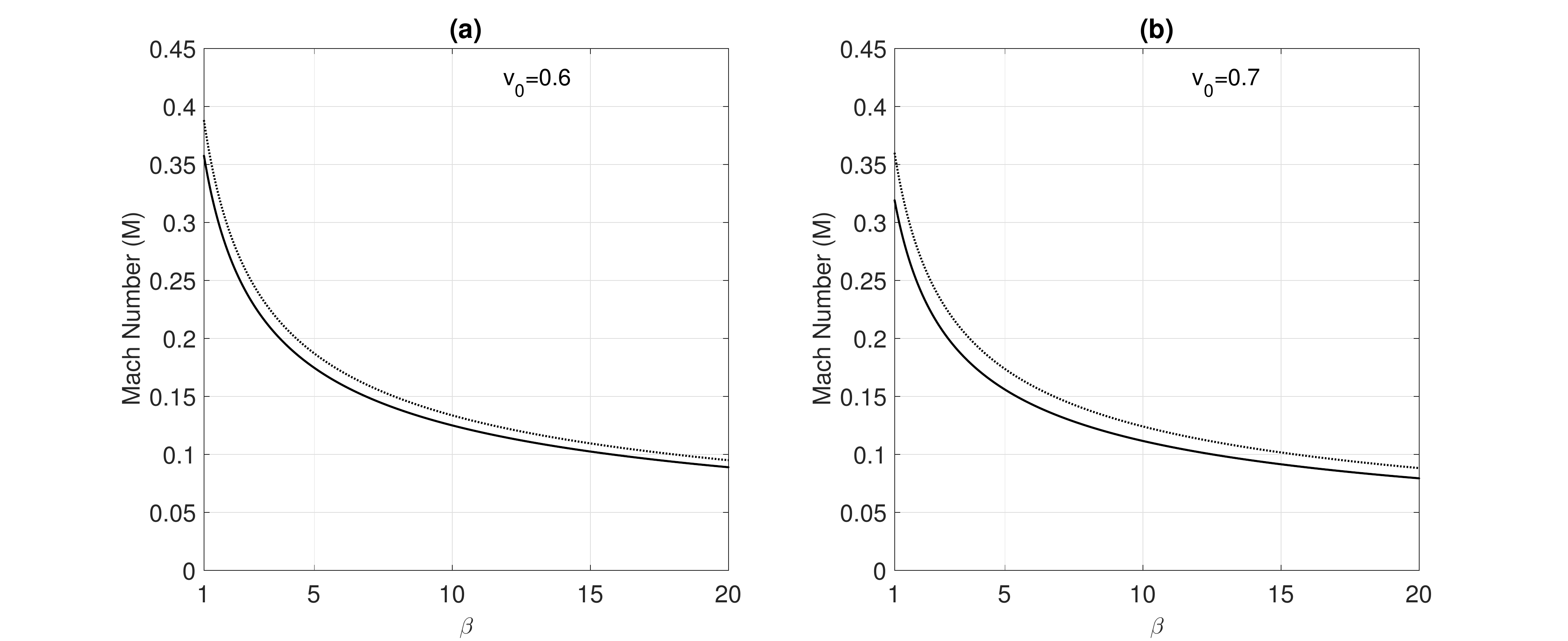}
\caption{Plots of the lower $(M_c)$ and upper $(M_u)$ limits of the Mach number, given by Eqs. \eqref{eq-Mc} and \eqref{eq-Mu}, are shown   for different values of $v_0$ $(\sqrt{1/5}<v_0<1)$  in ultrarelativistic plasmas $(\beta \gg 1)$: (a) $v_0=0.6$ and (b) $v_0=0.7$.  }
\label{fig:McMu_ultra}
\end{figure*}
\par
Figure \ref{fig:McMu_ultra} shows the plots of $M_c$ (the lower limit of the Mach number, solid line) and $M_u$ (the upper limit of the Mach number, dashed line)   within the domain $\sqrt{1/5}<v_0<1$  for different values of $v_0$.    We find that  both $M_c$ and $M_u$ decrease with increasing values of $\beta$ and they remain less than unity even for $\beta\gg1$. Furthermore, it is noticed that the values of both $M_c$ and $M_u$ decrease with increasing values of $v_0$.   Thus, it follows that in contrast to the weakly relativistic regime,   the EP plasmas with ultrarelativistic energies may support only sub-Alfv{\'e}nic solitons. Such a feature in relativistic EP plasmas has not been reported before.     
\par
\begin{figure}[h!]
\centering
\includegraphics[scale=0.30]{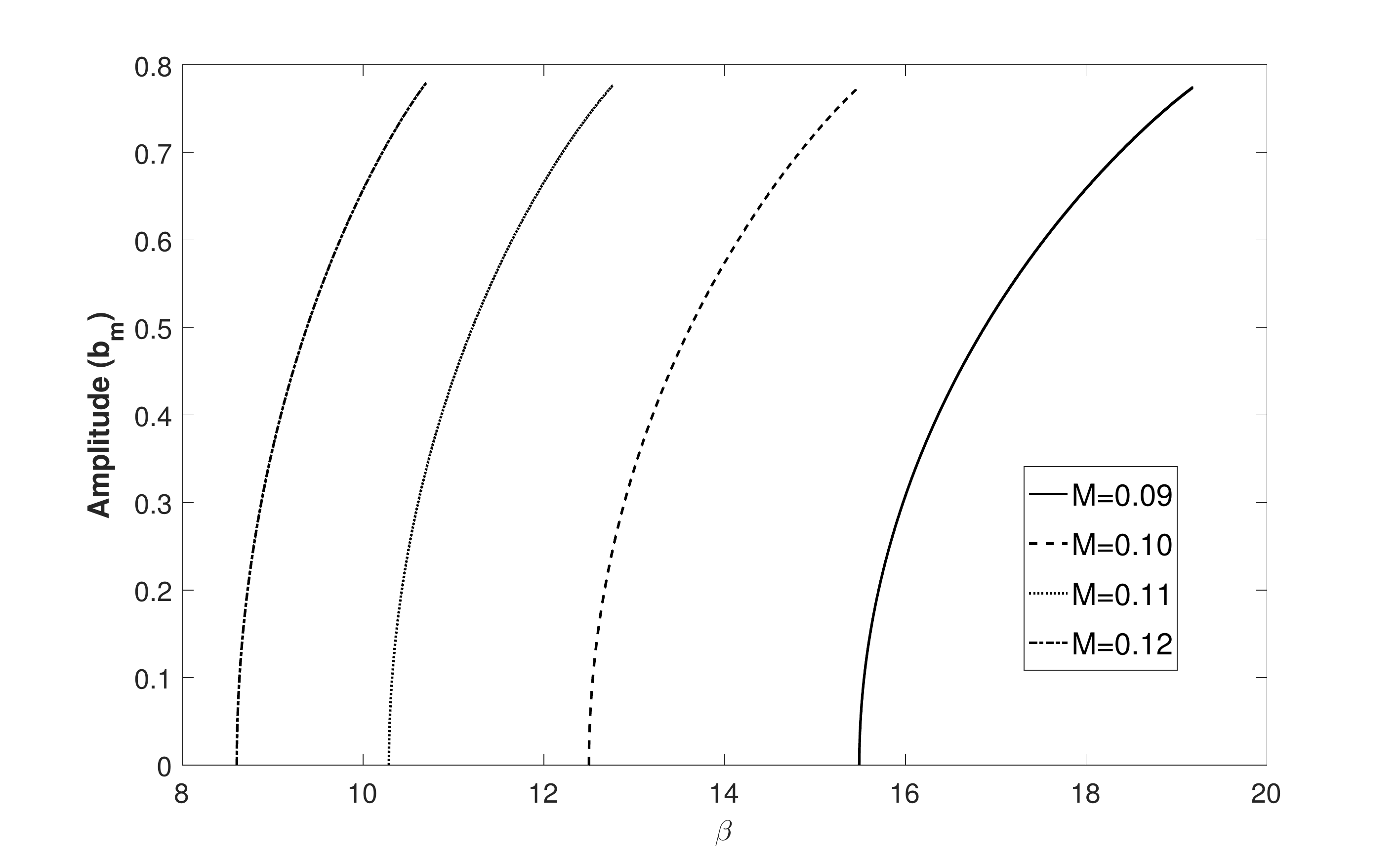}
\caption{The soliton amplitude $b_m$  is shown against $\beta$ in the ultrarealtivistic regime $(\beta\gg1)$ for different values of the Mach number $M$ and for a fixed value of $v_0=0.7$   in  $\sqrt{1/5}\leq v_0<1$.}
\label{fig:amp_beta_ultra_v07}
\end{figure}

  
Similar to the case of weakly relativistic plasmas, we also show the variation of the soliton amplitude $b_m$ for different values of the Mach number  $M$ within $M_c<M<M_u$ and with a fixed value of $v_0$ in $\sqrt{1/5}<v_0<1$  as shown in Fig. \ref{fig:amp_beta_ultra_v07}. It is found that  the values of $b_m$ increase  with increasing values of $\beta$, however, the threshold values  of $\beta$  shift  to   lower ones  as the values  of $M$ are increased. Since $\beta\gg1$, relatively lower values of $M$ would favor the existence of EM solitary waves in ultrarelativistic regimes.    
\par
\begin{figure*}
\includegraphics[scale=0.36]{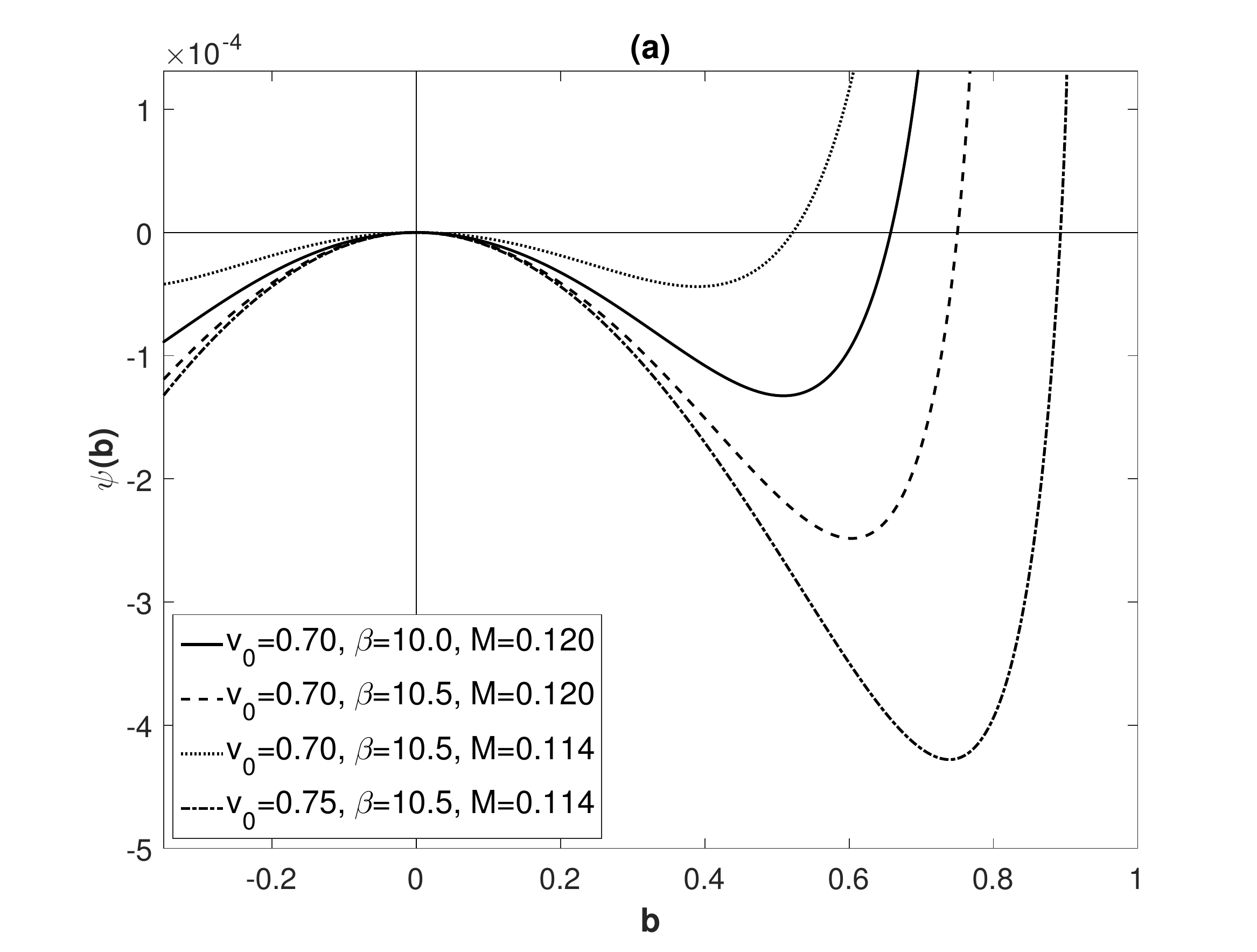}
\includegraphics[scale=0.36]{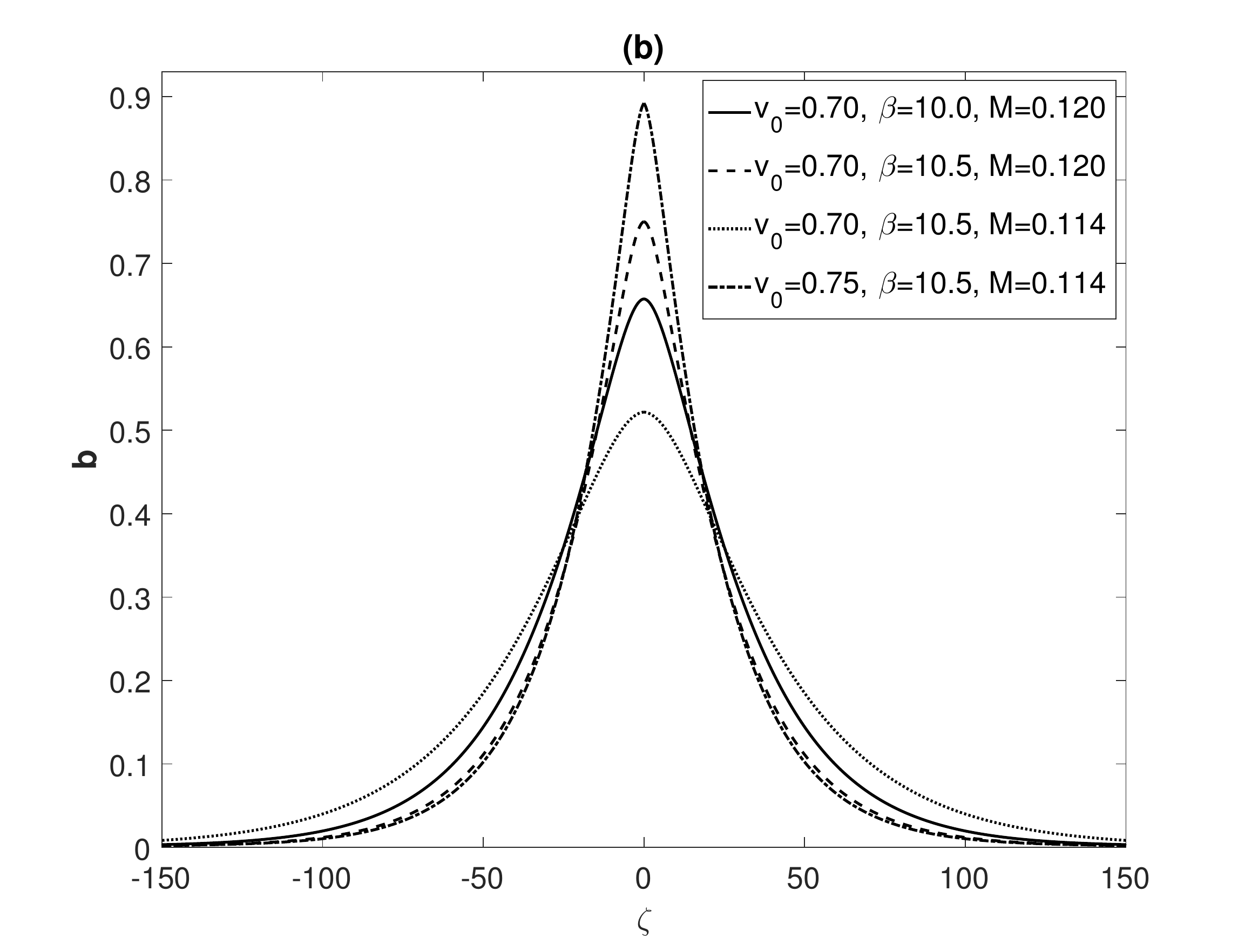}
\caption{Profiles of the pseudopotential $\psi(b)$ [subplot (a)] and the corresponding soliton [subplot (b)]  are shown in ultrarelativistic ($\beta\gg1$) regime for different values of $v_0$, $\beta$ and $M$ as in the legends with $\sqrt{1/5}\leq v_0<1$ and $M_c<M<M_u$.}
\label{fig:psi_ultra_v07}
\end{figure*}

 
  The pseudopotential $\psi(b)$ and the corresponding soliton profiles of the magnetic field $b$  are also shown in Fig. \ref{fig:psi_ultra_v07}   for different values of $v_0$, $\beta$ and the Mach number $M$.   It is   seen that with increasing values of these parameters, the soliton amplitude  increases and the width  decreases. 
\section{Conclusion}
 We have studied the nonlinear propagation of  purely stationary large amplitude electromagnetic solitary waves in a magnetized relativistic electron-positron-pair plasma. A fully relativistic two-fluid model is considered which accounts for both the weakly relativistic $(\beta\ll 1)$ and ultrarelativistic $(\beta\gg 1)$  thermal motions of electrons and positrons where $\beta\equiv k_BT/mc^2$. Thus, previous theory in the literature \citep{verheest2004} is advanced and generalized. Using the McKenzie approach, the system of fluid equations is reduced to an energy-like equation which describes the evolution of EM solitary waves in its own reference frame.  Different  parameter regimes of the wave phase velocity $v_0\equiv V/c$ and the energy ratio $\beta$ for the existence of  solitary waves, as well as different limits of the soliton amplitude $(b_m)$ and the Mach number $M\equiv V/V_A$   are demonstrated both in  the limits of weakly relativistic and ultrarealtivistic energies. It is found that 
 \begin{itemize}
\item In the weakly relativistic limit, EM solitary waves may exist  in two different regimes (i) $0<v_0<\sqrt{2/9}$, $0<\beta<v_0^{2}/(1-7v_0^{2}/2)$   and (ii) $\sqrt{2/9}\leq v_0<1$, $0<\beta \ll 1$. The solitary waves can appear as  the sub-Alfv{\'e}nic $(0<M<1)$ or super-Alfv{\'e}nic $(1<M<\sqrt{2})$ solitons   with amplitude $0<b_m<2$. 
\item In the ultrarelativistic limit, EM solitary waves exist  in the regime    $\sqrt{1/5}<v_0<1$, $\beta\gg1$. In this case, only sub-Alfv{\'e}nic $(0<M<0.4)$ solitons may exist with amplitude $0<b_m<1$. 
\end{itemize} 
\par 
It is to be noted that both the sub-Alfv{\'e}nic and super-Alfv{\'e}nic  solitons exist symmetrically for the wave magnetic field $b\equiv B_y/B_0>0$ or $<0$ owing to the obvious symmetry of EP-pair plasmas with equal   mass and opposite charges. This means that the EM solitary waves can propagate as compressive or rarefactive  type solitons.  The energy integral is expressed in terms of the magnetic field instead of the electrostatic potential as the latter may be relevant for electrostatic solitary waves not for EM waves. Furthermore, we have considered the isothermal pressure law for mathematical simplicity.  Instead, one can  use the adiabatic pressure law, i.e.,   $P/P_0=(n/n_0)^{\Gamma}$ with polytropic index $\Gamma$, however, in this case, the relativistic fluid equations may not be reducible to the energy integral form \eqref{eq:EnergyIntegral} either by the McKenzie approach or Sagdeev approach. 
\par 
To conclude, the nonlinear excitation of  EM waves and the formation of solitary structures in pair plasmas are known to have significant relevance not only in space and astrophysical    environments but also in laboratory experiments \citep{sarri2015}. Furthermore, in pulsars and active galactic nuclei with violent surroundings, these nonlinear phenomena would not occur with small amplitude only. In this context, the present theory in magnetized electron-positron plasmas can help understand certain aspects of these stronger nonlinear phenomena with large wave amplitude.
\section*{Acknowledgments}
This work was initiated when Sayantan Dutta was pursuing his Master's degree final project in the Department of Mathematics of Visva-Bharati.   One of us, GB acknowledges financial support from University Grants Commission (UGC), Govt. of India, under the Dr. D. S. Kothari Post Doctoral Fellowship Scheme with Ref. no. F.4-2/2006(BSR)/MA/18-19/0096). APM is supported by  the Science and Engineering Research Board (SERB), Govt. of India with  Sanction  order no. CRG/2018/004475   dated 26 March 2019. 

\bibliography{ManuscriptJune20}
\bibliographystyle{elsarticle-harv}
\end{document}